 \documentclass[final,3p,times,twocolumn]{elsarticle}

\usepackage{amssymb}
\usepackage{amsmath}
\usepackage{color}

\journal{Physica E: Low-dimensional Systems and Nanostructures}

\begin{document}

\begin{frontmatter}



\title{Interplay Between Stacking Order and In-plane Strain on the Electrical Properties of Bilayer Antimonene}

\author{Shoeib Babaee Touski}
\address{Department of Electrical Engineering, Hamedan University of Technology, Hamedan, Iran}
\author{Nayereh Ghobadi}\ead{n.ghobadi@znu.ac.ir}
\address{Department of Electrical Engineering, University of Zanjan, Zanjan, Iran}

\begin{abstract}
In this work, the electrical properties of bilayer Antimonene with different stacking orders are studied. Density functional theory with van der Waals (vdW) correction is used to investigate the electrical performances. Two configurations demonstrate considerable bandgaps, whereas, the bandgaps are close to zero for other structures. The in-plane biaxial strain is applied to modify the electrical properties. The bandgap reaches a maximum at a specific strain level and then closes at more enormous compressive and tensile strains. The energy of three valleys ($\Gamma$, Q, and K) in the conduction band are explored with the strain. The conduction band minimum switches between these valleys with the strain. Two bands also contribute to the valence band maximum, and the energy of these two bands for various strains is investigated. Finally, the effective mass for the valleys of the conduction band and the valence band are obtained. The effective mass at $\Gamma$-valley demonstrates the lowest effective mass.  
\end{abstract}


\end{frontmatter}


\section{introduction}
Recently, the two-dimensional family of group-VA (P, As, Sb, Bi) has attracted a high interest for its unique properties. These materials exhibit high stability, high carrier mobility, and good thermal conductivity \cite{wang2015atomically,zhang2016semi,zhang2018recent,ares2018recent}. These properties, along with high electrical conductivity, make them a promising candidate for future electronic\cite{pizzi2016performance}. The group-VA species, such as arsenic (As) \cite{yu2018tunable}, antimony (Sb) \cite{zhang2015atomically}, and bismuth (Bi) \cite{pumera20172d} monolayers with buckled nanostructures are found to be more stable.

Antimonene (Sb monolayer), a member of this family, has attracted huge interest since its stability was reported by Zhang, et al. \cite{zhang2015atomically}. They predicted that the rhombohedral phase ($\beta$-phase) has the highest stability between several possible structures of antimonene. Sb monolayer with $\beta$-phase has been studied experimentally and theoretically. This phase consists of Sb atoms buckled in a honeycomb lattice and exhibits an indirect bandgap \cite{ares2016mechanical,wang2015atomically}. This monolayer demonstrates high stability, high carrier mobility for both electron and hole, and good thermal conductivity \cite{ji2016two,pizzi2016performance}. These properties make antimonene as a promising candidate for future electronic devices. This monolayer displays a buckled honeycomb structure with D$_{3d}$ point group. Monolayer antimonene can be obtained by different methods, including micromechanical exfoliation \cite{ares2016mechanical}, liquid-phase exfoliation \cite{gibaja2016few}, electrochemical exfoliation \cite{lu2017broadband}, van der Waals epitaxy \cite{ji2016two} and molecular beam epitaxy \cite{chen2018single,wu2017epitaxial}.

Instability of 2D materials is one of the challenges for their application. Since the stability of antimonene in air and water has been approved by experiment and simulation \cite{wu2017epitaxial,ares2016mechanical}, antimonene is introduced as a promising candidate for various applications \cite{morishita2015,kuriakose2018black}. Monolayer Sb is an indirect band semiconductor with a theoretical bandgap in the range of 1.2-2.38 eV \cite{singh2016antimonene,wang2017many}.

\begin{figure*}
	\centering
	\includegraphics[width=1.0\linewidth]{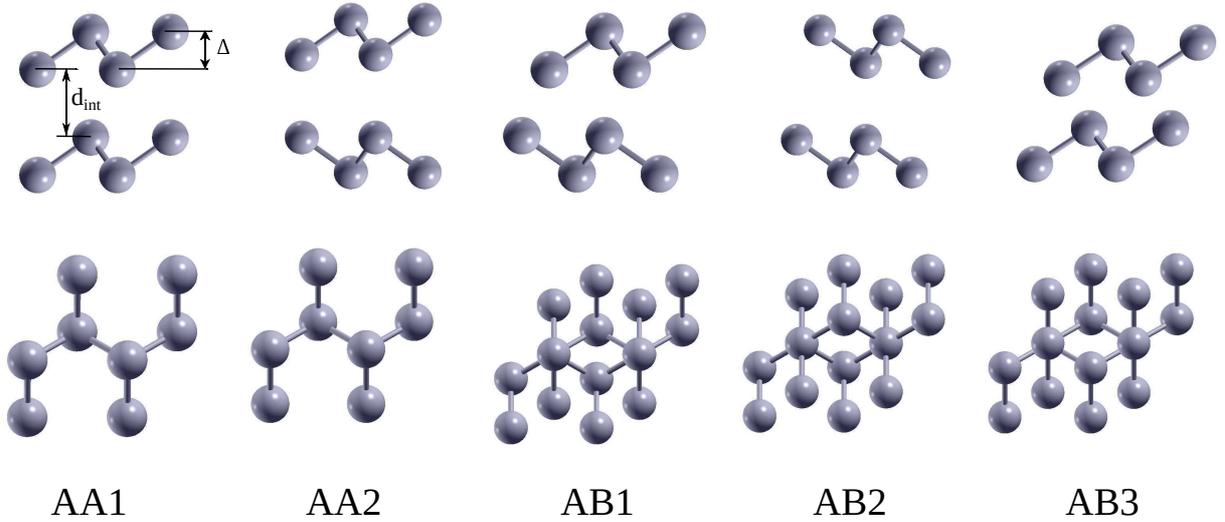}
	\caption{Structures of five different stacking order of bilayer antimonene.}
	\label{fig:schem}
\end{figure*}

Antimonene exhibits a high charge carrier mobility comparable to the other 2D semiconductors \cite{pizzi2016performance,wang2017many,chang2018novel}. In addition to the electronic application \cite{martinez2018antimonene,sun2018sub}, these properties make antimonene suitable for application in optical devices \cite{singh2016antimonene}. Several studies demonstrate that antimonene is a promising alternative material for optoelectronic applications where the strain-tunable bandgap could provide additional control over the material’s properties \cite{zhao2015strain,kripalani2018}.

The high-quality multi-layer antimonene has been successfully synthesized by some works \cite{tsai2016advent,ji2016two,lu2017broadband}. Multilayer antimonene nanoribbons are successfully synthesized by the plasma-assisted process and demonstrate a bandgap of 2.03 eV with RT orange light emission \cite{tsai2016advent}. Based on the successful synthesis of mono- and multi-layer antimonene, its bilayer will be reported in the future. There are many works that have studied the monolayer antimonene, whereas, its bilayer is not explored considerably. Monolayer antimonene has been reported as an indirect semiconductor with a considerable bandgap, whereas, an abrupt transition from semiconductor to metal takes place for its bilayer \cite{zhang2015atomically,akturk2015single}. Meiqiu Xie, et al. \cite{xie2017van} have studied three different bilayer stacking orders. They reported that these three configurations can exhibit bandgaps from 0.09 to 0.62 eV based on their staking orders. They demonstrated that AB stacking has the lowest energy and highest stability and displays the smallest bandgap. Two years later, Xiaoxu Wang, et al. \cite{wang2019good} reported two different AA-stacking and AB-stacking with the bandgaps of 0.35 eV and 0.04 eV, respectively. There are different stacking orders have not been studied. 

In this work, five stacking orders of bilayer antimonene are studied and their electrical properties are compared. Then biaxial strain has been applied to these five configurations to modify the electrical properties. 

\begin{figure*}
	\centering
	\includegraphics[width=0.8\linewidth]{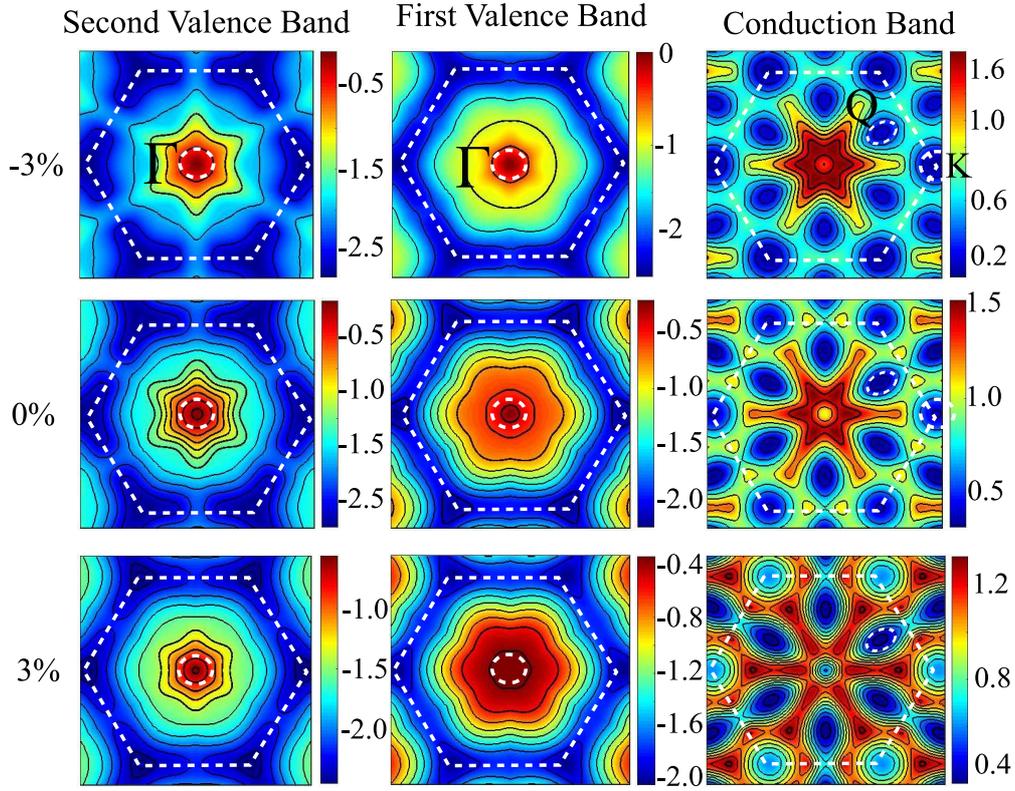}
	\caption{ The first and second valence and conduction band energy in the first Brillouin zone of AA2 structure. The figure is repeated for -3$\%$, 0 and 3$\%$ strains. $\Gamma$ valley is highlighted for the first and second valence bands and K and Q valleys for the conduction band.}
	\label{fig:fig2}
\end{figure*}

\begin{figure*}
	\centering
	\includegraphics[width=1.0\linewidth]{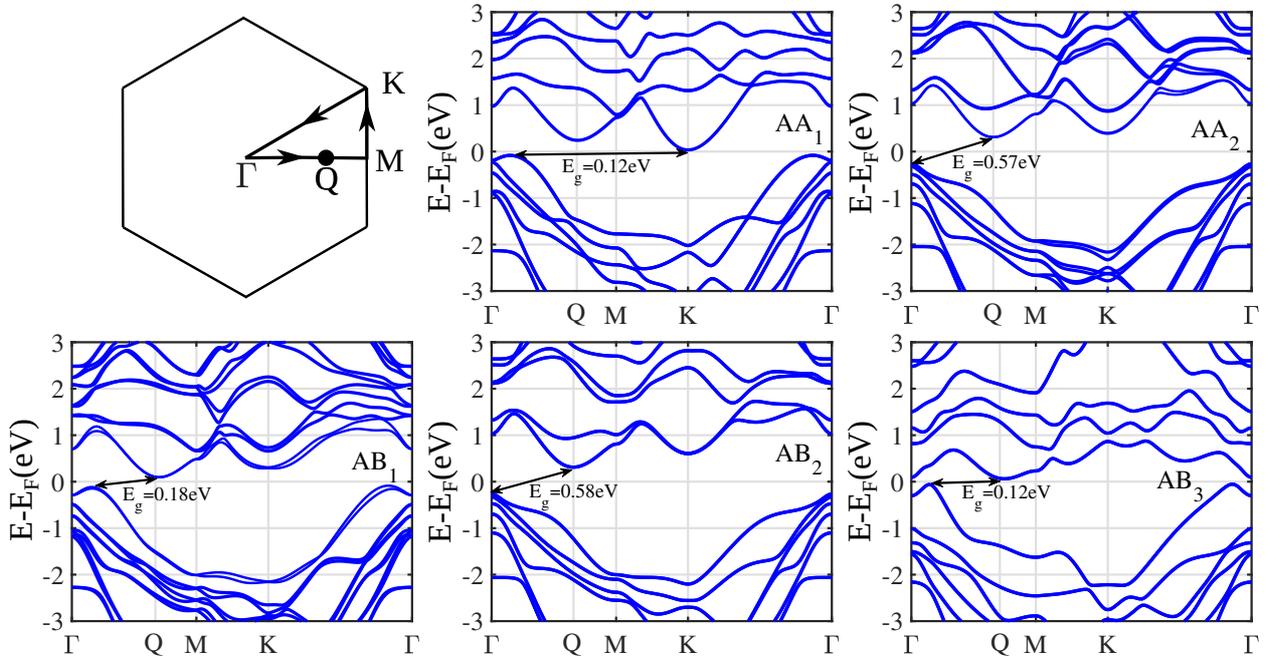}
	\caption{Band structures of bilayer antimonene with different stacking orders.}
	\label{fig:fig3}
\end{figure*}

\section{Computational details}
In order to investigate the different stacking orders of bilayer antimonene, density functional calculations are performed using the SIESTA package \cite{soler2002siesta}. The Perdew-Burke-Ernzerhof (PBE) \cite{perdew1981self} exchange-correlation functional of the generalized gradient approximation (GGA) has been employed. The van der Waals (vdW) interaction between adjacent layers has been treated with the Grimme's (DFT-D2) correction to the PBE functional in SIESTA \cite{grimme2006}. DFT calculations with vdW interaction are accomplished using the same method described in our previous study \cite{ghobadi2019normal}. Spin-orbit coupling (SOC), which has a significant effect on heavy elements such as antimony, is also included in our calculations. A double-ζ$\zeta$ plus polarization basis-set is used, and the energy cutoff is set to be 50 Ry. The total energy is converged to better than $10^{-5}$ eV, and the structures are fully relaxed until the force on each atom is less than 0.01 eV/$\AA$. A Monkhorst-Pack Brillouin-zone k-point grids of $21\times21\times1$ is chosen for the unit-cell of bilayer antimonene. The atomic structure is modeled by a periodic boundary condition, and a 30$\AA$ thick vacuum layer is used to eliminate the interlayer interaction in the normal direction. In order to visualize the atomic structures, the XCrySDen package has been employed \cite{kokalj2003}. A biaxial in-plane strain has been applied to the different stacking orders which are defined as $\epsilon=(a-a_0)/a_0$,
where $a_0$ and $a$ are the equilibrium and deformed lattice constants, respectively. The effective masses of the electrons and holes are calculated as \cite{touski2020electrical}:
\begin{equation}
m^*=\hbar^2/\left(\partial^2E/\partial k^2\right)
\end{equation}
where $\hbar$ is reduced Planck constant, $E$ and $k$ are the energy and wave vector of the conduction band minimum and the valence band maximum.

\section{results and discussion}
Five different stacking orders of bilayer antimonene are shown in Fig. \ref{fig:schem}. Two configurations that top layer exactly placed on the underlying layer has been named AA1 and AA2. AA1 and AA2 own mirror and inversion symmetries, respectively. In the other three structures, the top layer has been shifted by one bonding length relative to the underlying layer. AB2 and AB3 have inversion symmetry whereas, AB1 doesn't show any inversion symmetry. The binding energies ($E_b$) of the structures are reported in Table \ref{tab:tab1}. $E_b$ is defined as $E_b=E_{BL}-2E_{ML}$, where $E_{ML}$ and $E_{BL}$ are the total energy of the relaxed monolayer and bilayer antimonene, respectively. AB3 demonstrates the lowest binding energy and is the most stable configuration. This configuration also has the highest lattice constant, the lowest interlayer distance and buckling height, and the smallest bandgap. Two configurations AB1 and AA1 have the lowest binding energy and the lowest interlayer distance after AB3. These two structures also display a bandgap close to zero. Two configuration AA2 and AB2 indicate a considerable bandgap whereas, they have the highest binding energy. These two structures also demonstrate the lowest lattice constant and the largest interlayer distance and buckling height. Due to the high interlayer distance, AA2 and AB2 behave as two separate monolayers and preserve the monolayer properties. One can conclude that there is a relation between the binding energy, bandgap, interlayer distance, buckling height, and the lattice constant. The structures with lower binding energy have a lower bandgap, smaller interlayer distance, smaller buckling height, and higher lattice constant.

\begin{table}[]
	\caption{The lattice constant ($a$), interlayer distance (d$_{int}$), buckling height ($\Delta$), binding energy ($E_b$) and bandgap ($E_g$) of five different stacking orders of bilayer antimonene.}
	\begin{tabular}{p{0.8cm}p{0.8cm}p{0.9cm}p{0.9cm}p{1.0cm}p{0.9cm}}
		\hline\hline
		& a($\AA$) & d$_{int}(\AA)$ & $\Delta(\AA)$ & $E_b(eV)$ & $E_g$(eV)  \\ 
		\hline
		AA1 & 4.071  & 3.028  & 1.670  & -1.416  & 0.117   \\
		AA2 & 4.053  & 3.978  & 1.680  & -1.156  & 0.574   \\  
		AB1 & 4.091  & 2.854  & 1.652  & -1.444  & 0.181   \\  
		AB2 & 4.054  & 3.919  & 1.680  & -1.163  & 0.577   \\  
		AB3 & 4.177  & 2.225  & 1.600  & -1.490  & 0.115   \\  
		\hline \hline
	\end{tabular}
	\label{tab:tab1}
\end{table}

\begin{figure}
	\centering
	\includegraphics[width=1.0\linewidth]{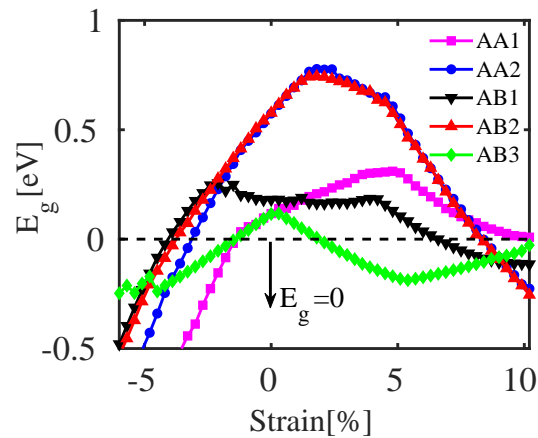}
	\caption{bandgap as a function of strain for different stacking orders.}
	\label{fig:fig4}
\end{figure}

The first and second valence bands and the conduction band of AA2 structure are plotted in Fig. \ref{fig:fig2}. In addition, these plots are repeated for three samples with compressive, relaxed, and tensile strains. As one can observe, $\Gamma$ point is the top of the first and second valence bands. On the other hand, two valleys are observed in the conduction band, K-valley and a valley that are located in a path from M to $\Gamma$-point and is called Q-valley. There exist six Q-valleys and two K-valleys in the first Brillouin zone.  Q- and K-valleys contribute to the conduction band minimum (CBM) for equilibrium and compressive samples, whereas, two $\Gamma$- and Q-valleys contribute to CBM for tensile strains.

The Band structures of five different stacking orders are reported in Fig. \ref{fig:fig3}. One can observe that all stacking orders demonstrate an indirect bandgap. The Conduction band minimum is located at Q-point for all structures except structure AA1 that CBM is located at K-point. On the other hand, the valence band maximum (VBM) is located at $\Gamma$-point for structures AA2 and AB2 whereas, the VBM is located at $\Gamma^*$-point close to $\Gamma$-point for other structures. The values of the bandgaps are written in the middle of these plots. The bandgaps are close to zero for three structures (AA1, AB1, and AB3), and only two structures (AA2 and AB2) have considerable bandgaps. With looking at AA2 and AB2 structures, one can find that one atom from the top layer is exactly placed on the one atom of the underlying layer with the minimum distance. The calculated bandgaps are 0.57 and 0.58 eV for AA2 and AB2 stacking orders that are close to the 0.62 eV reported by previous work \cite{xie2017van}. 

Sb atom is a heavy element, and spin-orbit coupling has a considerable effect on its electrical properties and bandgap modification \cite{rudenko2017}. Therefore, SOC is turned on for all configurations. Spin splitting does not occur in the structures AA1, AB2, and AB3 that have inversion symmetry. Structure AA2 demonstrates spin splitting, whereas, there is not any spin splitting in the path from $\Gamma$ to M points due to its mirror symmetry \cite{ariapour2019spin}. Both CBM and VBM of this structure are located on this path.

The strain has been used as a common way to modify the electrical properties \cite{shamekhi2020band}. Here, we have applied biaxial strain to five structures, and their bandgaps are plotted versus strain in Fig. \ref{fig:fig4}. All configurations have a finite bandgap at equilibrium. The bandgaps reach a maximum value under specific strain values and then close at larger tensile and compressive strains. Therefore, a transition from semiconductor to semi-metal occurs at these strain values. The maximum values of the bandgaps and their corresponding strains, and the transition strains at compressive and tensile regimes are listed in Table \ref{tab:tab2}. 

\begin{figure}
	\centering	\includegraphics[width=1.0\linewidth]{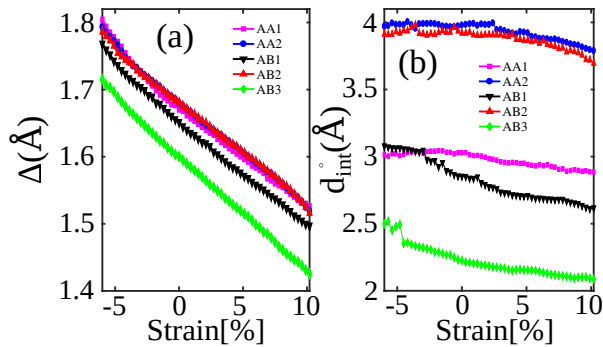}
	\caption{(a) Buckling heights and (b) interlayer distances versus strain for all stacking configurations.}
	\label{fig:fig5}
\end{figure}

\begin{figure*}
	\centering
	\includegraphics[width=0.98\linewidth]{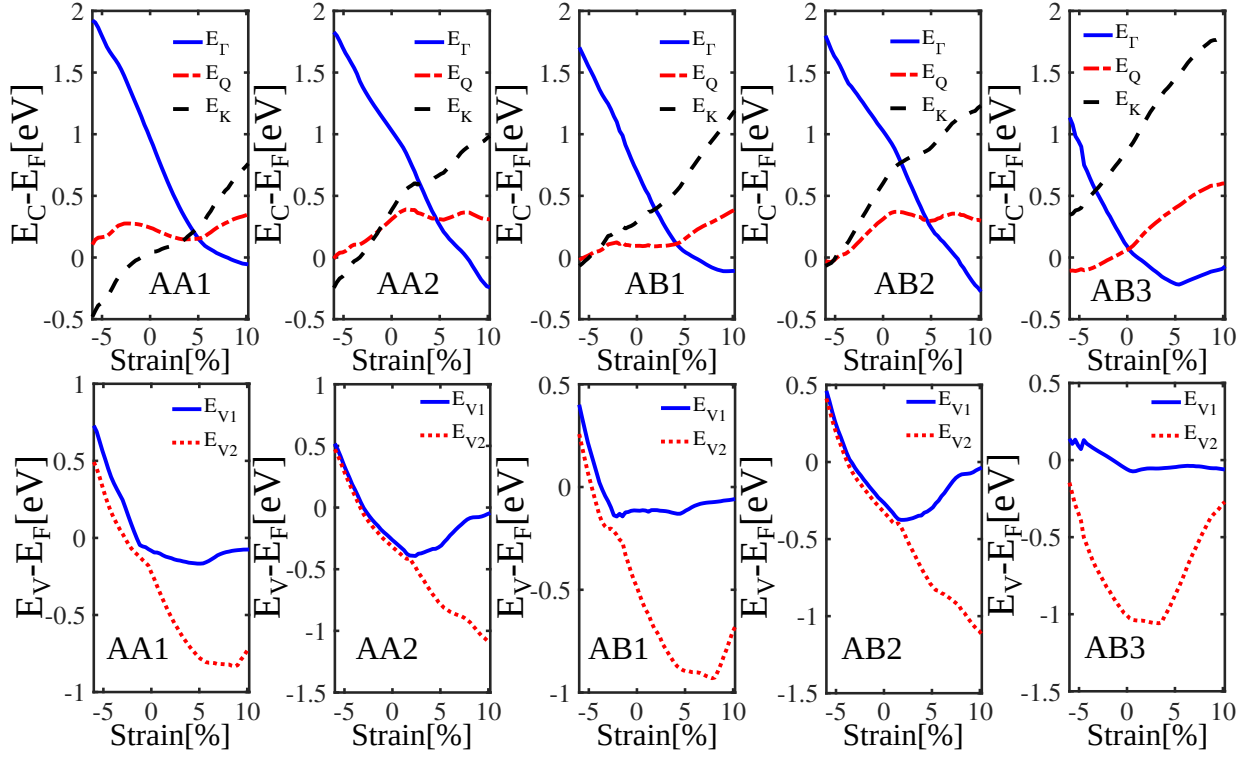}
	\caption{The energy of conduction band valleys at top row for five structures versus strains. The energy of first and second valence band maximum at bottom row.}
	\label{fig:fig6}
\end{figure*}

The AA1 stacking order has a bandgap of 0.117 eV at equilibrium. The tensile strain increases the bandgap, and a bandgap of 0.31 eV can be obtained at the strain value of 4.8$\%$. Further increase of the tensile strain reduces the bandgap, and a transition to semi-metal happens at 10.2$\%$ tensile strain. On the other hand, structure AA2 demonstrates a considerable bandgap, whereas, 1.8$\%$ tensile strain can raise the bandgap to 0.778 eV. Similar to AA1, the bandgap decreases with increasing strain up to 8.4$\%$, where the bandgap closes. The bandgap decreases with applying compressive strain in both structures, and bandgaps vanish at compressive strains of 1.5$\%$ and 3.3$\%$ for AA1 and AA2 stacking orders, respectively. 

The bandgaps of the AB series are also reported in Fig. \ref{fig:fig4}. The bandgap of AB1 configuration decreases with applying tensile strain so that the bandgap reaches to zero at 6.6$\%$. In this configuration, small compressive strain increases the bandgap, and the maximum bandgap value of 0.256 eV takes place at -2.1$\%$ strain. After this strain value, the bandgap reduces, and the structure changes to semi-metal at -4.2$\%$ strain. The behavior of the bandgap of the AB2 structure is similar to AA2. The maximum bandgap of 0.774 eV happens at 1.8$\%$ tensile strain, and the bandgap closes at 3.9$\%$ compressive and 8.4$\%$ tensile strains. It can be concluded from Table \ref{tab:tab2} that the variations of the bandgap of AA2 and AB2 configurations with strain are similar except the transition to semi-metal at different compressive strain values. AB3 also demonstrates a maximum bandgap of 0.117 eV at 0.3$\%$ tensile strain. The bandgap closes with a little more tensile (2.1$\%$) and compressive (-1.5$\%$) strains. This configuration has a small bandgap over a narrow range of strain.

\begin{table}[]
	\caption{The semiconductor to semi-metal transition strain ($\epsilon_{trans}$) at compressive and tensile regimes. The maximum bandgap ($E_{g,max}$) along with corresponding strain ($\epsilon_{max}$).}
	\begin{tabular}{p{0.7cm} p{1.9cm} p{1.2cm}p{1.cm}p{1.cm}}
		\hline\hline
		& $\epsilon_{trans}(\%)$ Compressive & $\epsilon_{trans}(\%)$ tensile & $\epsilon_{max}(\%)$ & $E_{g,max}$(eV)  \\ 
		\hline
		AA1 & -1.5  & 10.2 & 4.8  & 0.310\\
		AA2 & -3.3  & 8.4  & 1.8  & 0.778 \\  
		AB1 & -4.2  & 6.6  & -2.1 & 0.256 \\  
		AB2 & -3.9  & 8.4  & 1.8  & 0.774 \\  
		AB3 & -1.5  & 2.1  & 0.3  & 0.117 \\  
		\hline \hline
	\end{tabular}
	\label{tab:tab2}
\end{table}

Variation of the interlayer distances and buckling heights of all stacking orders versus strain are plotted in Fig. \ref{fig:fig5}. As one can expect, interlayer distance decreases with increasing strain. Two most stable configurations (AB1 and AB3) exhibit more decline compare to the others. AA2 and AB2 configurations also vary similarly here. On the other hand, the buckling heights of the structures behave similarly and decrease with an increase in strain value. This indicates that buckling height doesn't depend on stacking orders. 

\begin{figure*}
	\centering
	\includegraphics[width=0.98\linewidth]{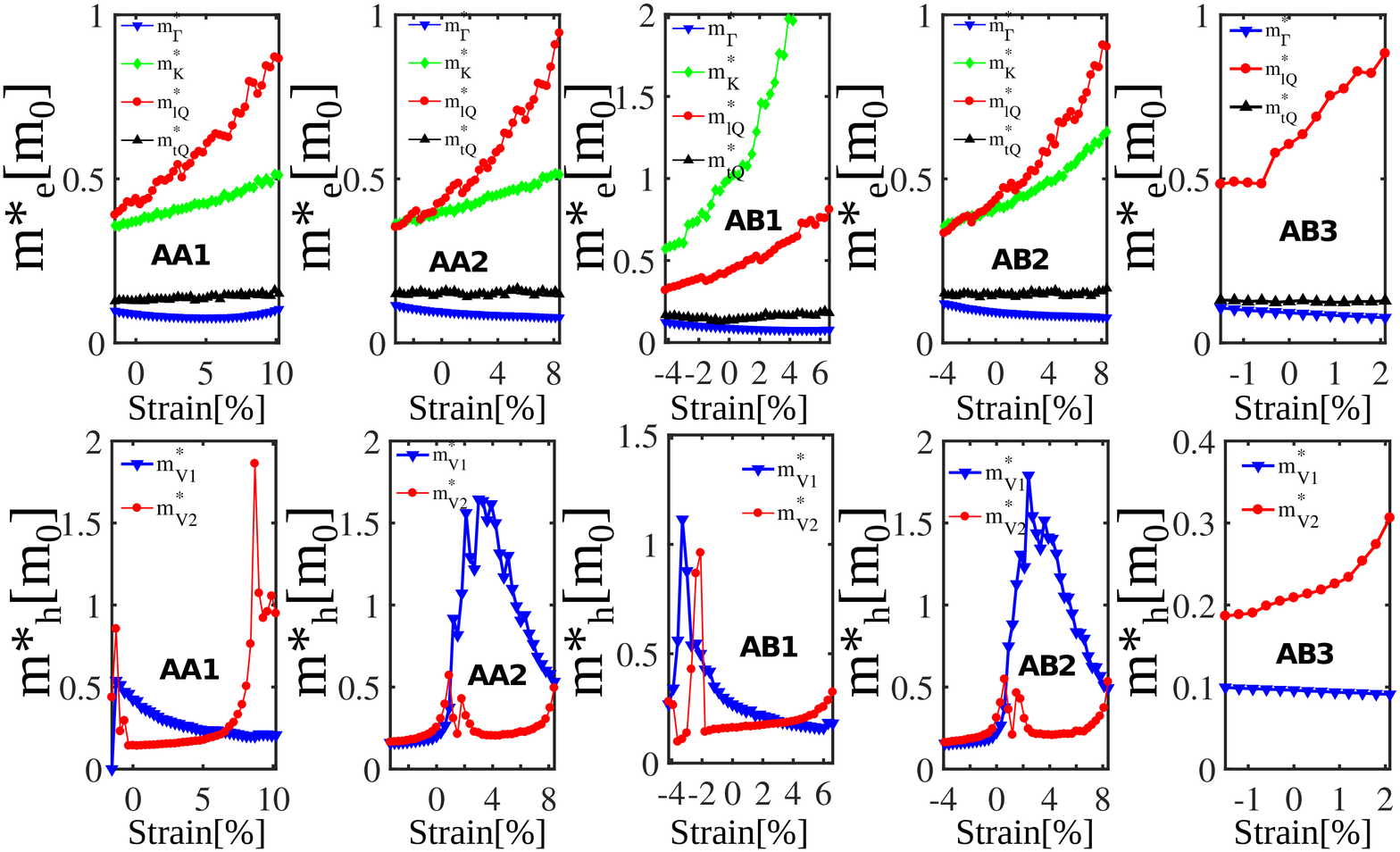}
	\caption{The effective mass of valleys in different configurations for electrons on the top row and for holes in the bottom row. The effective mass is reported for the strain values with a non-zero gap.}
	\label{fig:fig7}
\end{figure*}

Three valleys ($\Gamma$, Q, and K) in the conduction band are obvious from the band structures. The energy of these valleys for various values of strain is plotted in the top row of Fig. \ref{fig:fig6}. In all stacking orders, the energy of K-valley ($E_K$) increases with increasing strain whereas, the energy of $\Gamma$ ($E_\Gamma$) declines with strain. The AA1 structure is the only stacking with the CBM at K-valley in equilibrium. The CBM of this structure is located at K-valley in compressive strains. $E_K$ increases with applying tensile strain, and CBM changes to Q-valley at 3.6$\%$ tensile strain. The CBM remains at Q-valley in a small strain range of 3.6-5.1$\%$. $\Gamma$-valley is the conduction band minimum for strains larger than 5.1$\%$. In stacking AA2, the CBM is located at Q-valley in equilibrium. The energy of Q-valley reduces with applying compressive strain, but K-valley demonstrates more reduction and CBM changes to K-valley. The CBM remains at Q-valley for small tensile strain and then switches to $\Gamma$-valley in tensile strains larger than 4.8$\%$. In AB1 and AB2 stacking orders, the CBM stands at Q-valley under compressive strains. The CBM varies to K-valley at compressive strains of -4.5$\%$ and -5.1$\%$ in AB1 and AB2 configurations, respectively. The CBM remains at Q-valleys for small tensile strains in these two structures, whereas, the CBM switches to $\Gamma$-valley for tensile strains larger than 4.2$\%$ and 4.5$\%$ in structures AB1 and AB2, respectively. The AB3 stacking behaves differently so that the CBM is not located at K-valley for any strain. The CBM is placed at Q-valley and $\Gamma$-valley under compressive and tensile strains, respectively. 

Two bands contribute to the top of the valence band. The maximum energies of these bands as a function of strain are depicted at the bottom of Fig. \ref{fig:fig6}. The valence band maximum is located at $\Gamma$-valley in the AA2 and AB2 structures. The energy of these two bands is equal in these two structures and increases in compressive strains. Therefore, two bands contribute to the valence band maximum. The energy of the first band increases in tensile strain whereas, the second band decreases and goes farther from the first band. So, only the first band contributes to the VBM at the tensile strain regime. In three other structures, the maximum of the first band is located at $\Gamma^*$-point (a point near to $\Gamma$-point) whereas, the maximum of the second band is placed at $\Gamma$-point. A significant energy difference can be observed between these two bands in AA1 and AB1 stacking orders. The energy of the first and the second band increase in compressive strain. The energy of these two bands gets closer in compressive strain. The energy of the first band remains constant in tensile strain. On the other hand, the energy of the second band decreases first and then increases. The energy difference between these two bands is large in a tensile regime, so only the first band contributes to the VBM. This energy difference is large for all ranges of strain in stacking AB3. Therefore, only the first band is placed on the VBM.

In the following, effective masses of electrons and holes in different stacking orders are plotted in Fig. \ref{fig:fig7}. The effective masses are calculated for the strains with a non-zero bandgap. The electron effective mass is depicted in the upper row of the figure. The conduction band is constructed from three valleys ($\Gamma$, K, and M), and M-valley demonstrates two effective mass (longitudinal and transverse masses). The effective mass of electron at $\Gamma$-valley ($m^*_{e,\Gamma}$) shows the lowest effective mass. $m^*_{e,\Gamma}$ in all structures indicates a low dependency on the strain. In stacking AA1, $m^*_{e,\Gamma}$ decreases a little then increases for the larger strains. The lowest effective mass is $0.077m_0$ at the strain of 5.4$\%$, where the minimum conduction band is located at this valley under this strain. On the other hand, $m^*_{e,\Gamma}$ in four other structures (AA2, AB1, AB2, and AB3) decreases with increasing strain. The effective mass at K-valley ($m^*_{e,K}$) is depicted in four structures, and structure AB3 doesn't show any valley in K-point. $m^*_{e,K}$ increases with increasing strain. $m^*_{e,K}$ of AB1 structure shows a large value for tensile strains, whereas, CBM is located at Q- and $\Gamma$-valleys in these strains. Q-valley indicates two longitudinal and transverse effective masses. The longitudinal effective mass ($m^*_{e,lQ}$) is demonstrated as the highest effective mass in all structures except structure AB1. $m^*_{e,lQ}$ raises when strain increases. On the other hand, the transverse effective mass ($m^*_{e,tQ}$) is the second-lowest mass after $m^*_{e,\Gamma}$. $m^*_{e,tQ}$ is not influenced by strain and remains constant over all strains.

The bottom row of Fig. \ref{fig:fig7} indicates the effective mass of the hole in the first and second valence bands. The hole effective masses for two structures AA1 and AB1 behave similarly. The effective mass for the first band ($m^*_{v1}$) decreases with increasing strain whereas, the effective mass for the second band ($m^*_{v2}$) increases. Two structures AA2 and AB2 also behave similarly. Two effective masses $m^*_{v1}$ and $m^*_{v2}$ approximately remain constant in the compressive regime. In this strain regime, the first and second valence bands contribute to VBM. On the other hand, under tensile strain, the first band only is located at the VBM. Therefore, $m^*_{v1}$ is important in tensile regime. $m^*_{v1}$ raises for small strain and declines for larger strains. Structure AB3 behaves differently from other structures. The first band only contributes to the VBM, and $m^*_{v1}$ doesn't display any considerable dependency on the strain.  

\section{Conslusion}
In summary, the electrical and structural properties of five different stacking orders of bilayer antimonene are studied. Two configurations AA2 and AB2 demonstrate considerable bandgaps of 0.57 and 0.58 eV, respectively. Three other structures display a low bandgap of around 0.1 eV. The in-plane biaxial strain modifies the bandgap. The bandgap demonstrates a maximum at a specific strain level and then closes at larger tensile and compressive strains. The maximum values of the bandgaps are 0.778 eV and 0.774 eV for structures AA2 and AB2, respectively. The CBM switches between three valleys ($\Gamma$, Q, and K) under different strains. On the other hand, two bands contribute to the VBM at $\Gamma$-point. The energies of these two bands demonstrate that both of them are located at the VBM for compressive strains, whereas, one band contributes to the VBM in the tensile regime. Furthermore, the effective masses for three valleys of the conduction band are studied. The effective mass for $\Gamma$-valley displays the lowest effective mass for all range of strain, whereas, $m^*_{e,K}$ and $m^*_{e,Ql}$ demonstrate the highest effective mass. Finally, the hole effective masses for two bands are investigated over strains with a non-zero bandgap.

\bibliographystyle{elsarticle-num}
\bibliography{acronym,mx}


\end{document}